# Main Manuscript for

# Dual Effects of the US-China Trade War and COVID-19 on United States Imports: Transfer of China's industrial chain?

Wei Luo[1a]*, Siyuan Kang[1a], Sheng Hu[a], Lixian Su[a], Rui Dai[1b]*

[a] Geography Department, National University of Singapore, Singapore, 117568

[b] Wharton School of Business, University of Pennsylvania, US, PA 19104

*Corresponding Authors:

Wei Luo | Geography Department, National University of Singapore, Singapore, 117568 | Office Phone: (65) 6516-3851 | Email: geowl@nus.edu.sg

Rui Dai | Wharton School of Business, University of Pennsylvania, US, PA 19104 | Office Phone: +1(215)7464612 | Email: rdai@wharton.upenn.edu

## Abstract

The trade tension between the U.S. and China since 2018 has caused a steady decoupling of the world's two largest economies. The pandemic outbreak in 2020 complicated this process and had numerous unanticipated repercussions. This paper investigates how U.S. importers reacted to the trade war and worldwide lockdowns due to the COVID-19 pandemic. We examine the effects of the two incidents on U.S. imports separately and collectively, with various economic scopes. Our findings uncover intricate trading dynamics among the U.S., China, and Southeast Asia, through which businesses relocated portions of their global supply chain away from China to avoid high tariffs. Our analysis indicates that increased tariffs cause the U.S. to import less from China. Meanwhile, Southeast Asian exporters have integrated more into value chains centered on Chinese suppliers by participating more in assembling and completing products. However, the worldwide lockdowns over pandemic have reversed this trend as, over this period, the U.S. effectively imported more goods directly from China and indirectly through Southeast Asian exporters that imported from China.

---

[1] W.L., R.D., SY.K. contributed equally to this work.

# Introduction

In the realm of global trade, the United States and China, as the world's leading economies, have become embroiled in a protracted trade war since 2018, resulting in an unprecedented escalation of retaliatory tariffs. This contentious trade tension has significantly reshaped established worldwide supply-chain networks that have evolved over decades of international commerce, with enduring ramifications extending beyond the current year[1,2]. These mutually escalating tariffs, encompassing a staggering volume of approximately $600 billion in trade flows, have profoundly impacted bilateral trade across product categories over the global supply chain[3,4]. The intricate consequences of this trade war have been further compounded by the global outbreak of the COVID-19 pandemic, one of the most significant pandemics in human history. Existing literature, however, has primarily focused on the effects of the trade war or the pandemic mainly from the perspective of individual economies while overlooking the dynamic nature of the supply chain network over the two events. Fewer studies examining the combined impact of the two events primarily rely on theoretical calibration or simulation prediction. Namely, little is empirically known about the responses of economies with high spatiotemporal heterogeneity to those exceptional disruptions. Our study fills this void by applying innovative methodologies over a unique data set. We illustrate how the two events jointly reshaped the global supply chain networks till 2021, offer new insights into the economic consequences to various parties, and shed light on the future development of geoeconomics relationships.

Against the backdrop of these two far-reaching global events, the landscape of global trading and the global supply chain has become remarkably complex and sophisticated. Numerous studies have endeavored to examine the economic implications of the trade war and the COVID-19 pandemic individually[5–10]. The ongoing trade dispute between the United States and China has significantly influenced worldwide economic operations, leading to the restructuring of supply chains and generating extensive consequences throughout various industries, nations, and geographical areas. Regarding the domestic ramifications within the United States, industries associated with manufacturing and technology, specifically those with close ties to China, such as electronics, machinery, and automobiles, have encountered notable disruptions and escalated expenses[11,12] Some other studies have further explored the deeper impact of the US-China trade war on the pass-through of tariffs to importers and on welfare by including tariff-related variables in the model[4]. Moreover, the trade conflict has instigated a noteworthy reorganization of worldwide supply chains in which Southeast Asia such as Vietnam, Thailand, and Malaysia have emerged as the primary recipients[13]. Numerous corporations endeavored to broaden their product portfolio, resulting in a notable upswing in manufacturing operations within those Southeast Asian nations[14–17]. Several other nations and territories, including the European Union, Mexico, and South Korea, have endeavored to leverage the possibilities arising from supply chain relocation[18,19]. The reorganization of worldwide supply chains is exerting an impact on the intricacies of global commerce and manufacturing networks.

The reallocation of global supply chains due to the trade war has attracted scholars from numerous fields, including geography, logistics, management science, and economics. Economists have developed theoretical frameworks to comprehend the mechanism of value chain reallocation. Using various models, one body of economic literature suggests that tariffs could harm bilateral trade[13,20]; meanwhile, another demonstrates that exogenous shocks in product cost and market size would result in the reallocation of a firm's business activities[21–23]. Recent research[24] studies the reallocation effects caused by the U.S.-China trade war and suggests that, under typical circumstances, countries that substitute or complement Chinese or U.S. goods on the global supply chain benefit from the trade war [24]. Based on the presented premises, we expect that a trade war between the U.S. and China will inevitably lead to a decline in bilateral trade between the two countries. Furthermore, according to the gravity theory of trade [25–27], an economy will gravitate toward trading with its neighbors with similar cultural preferences and development stages. Therefore, we hypothesize that China's neighboring developing economies will be the initial beneficiaries of value chain allocations.

The intricate nature of the alterations in worldwide economic activity resulting from the abrupt emergence of the COVID-19 pandemic is noteworthy. Initially, the COVID-19 pandemic caused significant disturbances to worldwide supply chains, resulting in extensive manufacturing, transportation, and commercial exchange disruptions[28–30]. The implementation of trade embargoes, limitations on travel, and the cessation of manufacturing operations caused disruptions in the transportation of goods, resulting in impediments and scarcities within supply chains[7,31–35]. However, previous literature on epidemics and other supply chain disruptions provides us little theoretical guidance[36], making it difficult to postulate how the COVID-19 pandemic would affect the global supply chain allocation process. Only a handful of theories related to the pandemic disruption on the supply chain are from management disciplines. Those theories are usually founded on the propositions[32]: 1) a surged demand for essential products and 2) a significant contraction in raw material supply constraining production capacity over the pandemic period. Furthermore, global supply chains have experienced spatiotemporal heterogeneity of vulnerability because of the effectiveness of COVID-19 containment and the development and distribution of vaccines in different regions and countries[37,38]. China led Asia-Pacific economies to show remarkable resilience in securing the global supply chain because they efficiently contained the virus at an early stage[39,40]. Combining the theoretical components and observed facts, Chinese businesses retained a relatively stable labor supply and endured less from a lack of raw materials because of their considerably self-sufficient value chain system. Thus, we expect to observe a reverse flow of supply chains previously diverted from China due to the US-China trade war, especially in industries including healthcare, pharmaceuticals, and high technology [41]. Nonetheless, the extent to which the two significant incidents could have a net effect is an empirical question we will investigate in the current article.

Although some studies have attempted to examine the combined impact of the US-China trade war and COVID-19, most are either analytical or rely on model calibration or simulation predictions due to the inherent constraints of time and data availability[42,43]. Furthermore, most focus on individual countries and overlook the trading dynamic among various parties. In contrast, our research provides the first empirical study to assess two events' individual and combined impact by incorporating diverse economies and leveraging comprehensive data until 2021. By adopting a broader perspective encompassing global, continental, regional, and national dimensions, our study offers a comprehensive understanding compared to spatially focused analyses concentrating solely on specific regions or countries[44–46]. Moreover, we delve into the economic sectoral scale, investigating the common changes observed across all economic goods and the variations in how specific categories of goods are affected across different regions. Through a thorough analysis of US imports and the inclusion of an examination of Chinese exports to specific Asian countries, our research provides a more holistic and elucidating picture of how the triangular trade relationship between the US, China, and Southeast Asia has evolved because of these two significantly sequential events.

## Materials and Methods

### Materials

In our investigation, we employed the UN Comtrade database to explore the intricate realm of both US imports and China's exports. The UN Statistics Division collects, processes, and validates the data from approximately 200 countries, representing over 99% of the world's merchandise trade. We use the import or export records provided by the originating countries to ensure data accuracy, originality, and consistency in our analysis. In particular, for analyzing United States imports, we utilize trade records the US provides to UN Comtrade, which identify the United States as a reporter and other nations as its trading partners. Similarly, we analyze China's exports using data gathered by Chinese authorities.

We acquired an extensive compilation of monthly trade data, meticulously disaggregated at the 6-digit HS code level, facilitating a nuanced comprehension of distinct economic sectors. Specifically, the 6-digit HS code system identifies 5,439 categories of bilaterally traded products in our data. Following international trade literature[24], we further partition products into nine categories: agriculture, apparel, chemicals, materials, machinery, metals, minerals, transport, and miscellaneous (Table S2). Our dataset on US imports initially consisted of 6,403,989 trade records from 2015 to 2021. These trade records comprised data fields such as Trading Month, Reporting and Partner Country, Trade Direction, 6-digit HS Code, Good Description, and Trade Value (Figure S1). We conducted data integrity checks, including detecting missing data, assessing outliers, eliminating duplicate entries, and validating data by comparing it to the exact figures from US Census Bureau statistics at the monthly level. Finally, we aggregated the monthly trade data at the 9-category level, resulting in 14,781 transposed records (Figure S2). Through a similar process, we aggregated Chinese export data from 1,868,176 trade records that used 6-digit HS codes to 480 instances of 9 categories from 2016 to 2021 (Figure S3). The omission of 2015 data in Chinese analysis is due to the unavailability of Chinese export data in the UN Comtrade database.

## Methods

We employed innovative methodology to analyze the value chain reallocation by combining the power of iteratively multi-scale event studies and visualization. To conduct our event study analysis, we first utilize a difference-in-differences OLS regression with 'time' and 'country'/'region' fixed effects. This approach mitigates the omitted variable issues associated with unobservable heterogeneity and time-specific factors. The resulting regression coefficients were assessed at 10% or lower significance levels to determine the most appropriate one for different data pairs. Compared to other empirical works, such as the gravity model and those built directly from specific models, our approach is more robust and less sensitive to variable omissions, data outliers, and theoretical oversimplifications. Our iterative approach allowed for a comprehensive examination of economic sectors at both continental and national or regional levels, contributing to a deeper understanding of the nuanced dynamics and impacts of the events analyzed. Then, to present our massive results comprehensively, we rely on various visualization representations to illustrate the relative changes over numerous metrics.

Specifically, we present a multi-scale event study model to investigate the consequential impact of two pivotal events, the trade war and the COVID-19 pandemic, on U.S. import trade. Rooted in econometrics, the event study model offers a robust means to evaluate the effects of specific interventions by contrasting outcome changes over time between treatment and comparison groups[47]. Using longitudinal data from treatment and control groups with a quasi-experimental design, this model enables the establishment of a reliable counterfactual to quantify the magnitude of changes caused by exogenous shocks. As a desirable form of event analysis, it finds diverse applications across various domains, including economics, public health, and beyond.

Given the intricate interplay of the international landscape, we postulate that the event study model is amenable to estimating the impact of crucial events, such as the trade war and the COVID-19 pandemic, on global trade centered on U.S. imports and China's exports. We propose an innovative combination of the power of visualization and this well-established approach to illustrate otherwise hard-to-comprehend patterns from the unique U.S. trade data. We aim to discern the effects of the trade war and the COVID-19 pandemic on global trade dynamics at the continent and country levels.

By accounting for the trade value variable and incorporating geographically fixed effects and monthly fixed effects, our study seeks to provide credible estimations of the numerical consequences of these events on U.S. import trade and China's export trade. The model equation is as follows:

$$Trade_{(i,t)} = \alpha * Treat_i + \beta * Post + \gamma(Treat_i \times Post) + \theta_{(i)} + \delta_{(t)} + \varepsilon_{(i,t)}$$

where the U.S. import or Chinese export trade share value ($Trade_{(i,t)}$) is the explained variable. $Treat_i$ is a dummy variable that denotes a focal (treatment) region or country (set to 1) from the control regions or countries (set to 0). $Post$ is set to 1 for the trade war or COVID-19 pandemic period and 0 for the comparison or benchmark period. When we investigate the trade war effect or joint effect, we use the sample from 2015 to 2017 as a benchmark sample, while the sample from 2019 is the benchmark sample for the sole effect of the pandemic analysis. $Treat_i \times Post$ is a core explanatory variable, the interaction variable of $Treat_i$ and $Post$. $\theta_{(i)}$ and $\delta_{(t)}$ are country/region and year fixed effects, respectively.

Calculated by the model, the coefficient of this variable, that is $\gamma$, measures the effects of the trade war and the COVID-19 pandemic on U.S. import trade. Without fixed effect variables, $\gamma$ measures the differences between the change in the focal country's or region's trade share and that of benchmarking countries or regions after or before an event[48]. Region/country and monthly-fixed effects tend to subsume the treatment/post-dummy variables and deviate the $\gamma$ slightly away from the exact value from the difference-in-differences. Despite the deviation, social scientists still consider that the interaction coefficient from this regression can reasonably approximate the difference-in-differences in the sample.

## Results

### The U.S. Imports Analysis

Figure 1 depicts the trend of US imports over time, in which Figure 1a shows a counterintuitive tendency. From 2015 to 2021, the overall value of U.S. imports steadily increased until Oct 2018, despite intensified confrontation between the U.S. government and many of its trading partners besides China[13,49,50]. In March 2019, the imports from US trading partners other than China (ROW) began to rebound and eventually exceed its pre-trade war levels. The U.S. import value from China even continued to increase until Oct 2018 followed by a drop afterward, resulting in an average decline of -6.13 % by the end of 2019 (Table 1b). The COVID-19 outbreak precipitated a substantial reduction in imports from ROW and China. However, by June 2020, US imports from ROW started to grow and reached an all-time high by the conclusion of our sample period. Despite the intensifying trade war, the imports from China over the same period increased modestly by an average of 5.69% from phase 2 to 3 (Table 1b).

Figure 1b and Table 1 together reveal the impacts of the trade war and the pandemic across economic sectors. Imports from the Minerals sector increased by an average of 28.34% from Phase 1 to Phase 2, followed by Chemicals and Miscellaneous with average increases of 18.75% and 15.37%, respectively. After the emergence of COVID-19, the pandemic significantly and negatively affected Minerals and Transport imports from Phase 2 to Phase 3, with declines of -

25.34% and -12.42%, respectively. In contrast, over this period, imports of Materials, Chemicals, and Metals increased by an average of 18.27%, 17.53%, and 16.91%, respectively. These findings indicate that the impact of the trade war and the pandemic on international trade differ by geographic and economic sectors.

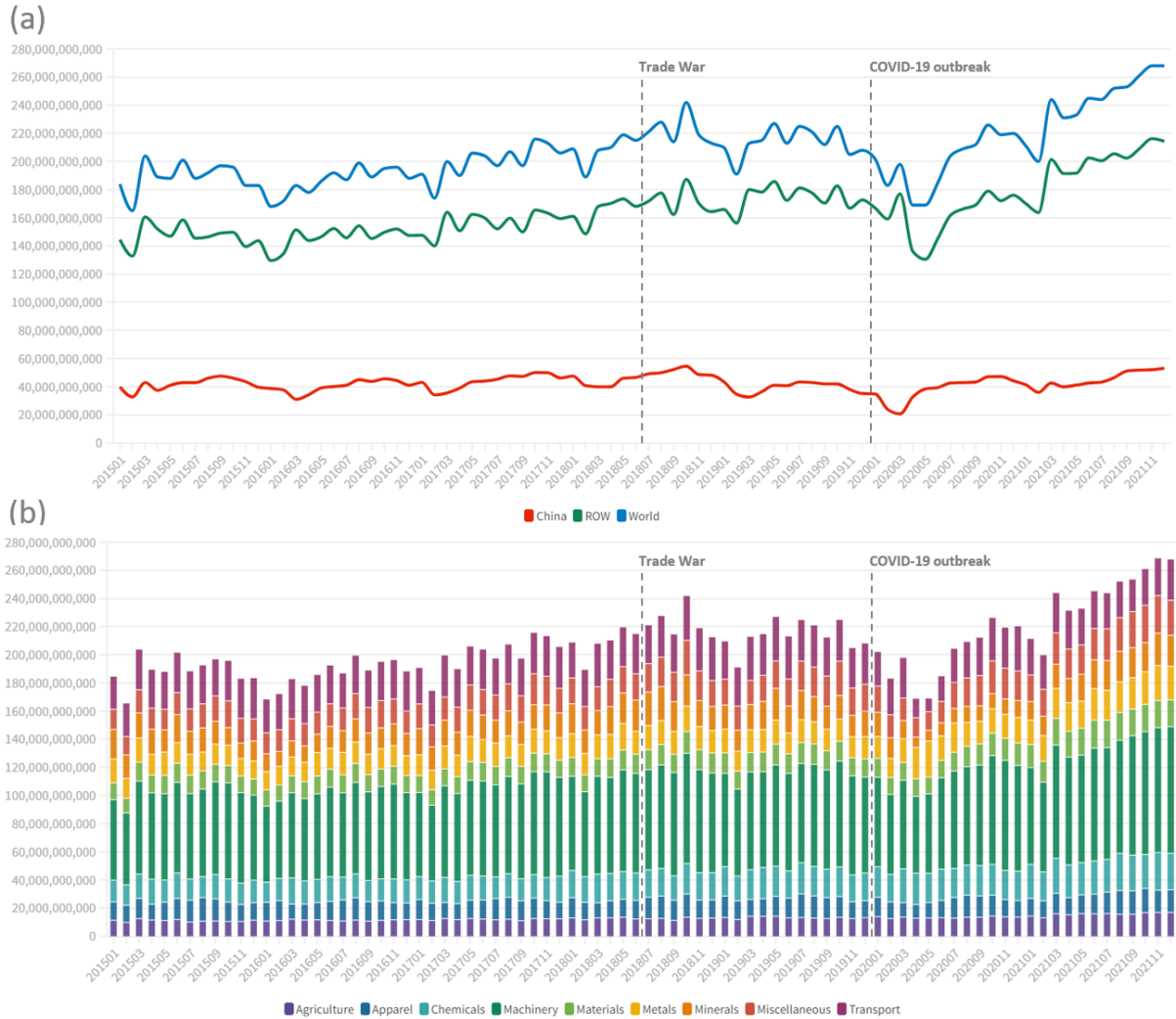

**Figure 1.** The monthly change of imports to the U.S. from 2015 to 2021. The unit of measurement is USD. Panel **(a)** illustrates the monthly U.S. imports from Mainland China and the rest of the world (ROW). Panel **(b)** illustrates the monthly U.S. imports across different economic sectors.

Table 1. The year-over-year (YoY) growth is the ratio of the difference between each period's average annual trade value and the average annual value of the previous period. Phase 1 includes three years in the pre-trade war period, from 2015 to 2017. Phase 2 is 2019, the year of the trade war that does not coincide with the Pandemic. Phase 3 lasts from 2020 to 2021, representing a period when the trade war and Pandemic overlap. This table contains (a) The Year-over-Year

(YoY) growth rates of the average total value of U.S. imports and (b) The Year-over-Year (YoY) growth rates of the U.S. import trade value from the top 20 countries or territories with the highest exports to the U.S.

**(a)**

| Sector | Examples | From Phase 1 to Phase 2 | From Phase 2 to Phase 3 |
|---|---|---|---|
| Agriculture | Soybeans, wine, coffee, beef | 10.31% | 15.05% |
| Apparel | Footwear, t-shirts, handbags | 1.57% | 0.87% |
| Chemicals | Medications, cosmetics, vaccines | 18.75% | 17.53% |
| Machinery | Engines, computers, cell phones | 11.85% | 3.54% |
| Materials | Plastics, lumber, stones, glass | 10.27% | 18.27% |
| Metals | Copper, steel, iron, aluminum | 12.12% | 16.91% |
| Minerals | Oil, coal, salt, electricity | 28.34% | -25.34% |
| Miscellaneous | Medical devices, furniture, art | 15.37% | 5.23% |
| Transport | Vehicles, airplanes, parts | 5.53% | -12.42% |

**(b)**

| Source | From Phase 1 to Phase 2 | From Phase 2 to Phase 3 |
| --- | --- | --- |
| Mainland, China | -6.13% | 5.69% |
| Mexico | 19.00% | -0.75% |
| Canada | 10.12% | -2.01% |
| Japan | 7.64% | -10.91% |
| Germany | 7.40% | -1.59% |
| Republic of Korea | 9.29% | 10.76% |
| Viet Nam | 57.88% | 37.93% |
| United Kingdom | 14.64% | -15.96% |
| Ireland | 38.52% | 12.78% |
| India | 24.03% | 8.97% |
| Taiwan | 32.40% | 27.95% |
| Italy | 23.03% | -2.93% |
| France | 19.78% | -18.31% |
| Switzerland | 29.54% | 53.09% |
| Malaysia | 13.16% | 24.32% |
| Thailand | 13.03% | 28.89% |
| Brazil | 11.84% | -10.75% |
| Netherlands | 75.88% | 4.35% |
| Singapore | 42.47% | 14.29% |
| Russia | 38.04% | 4.54% |

Figure 2 depicts the Year-over-Year (YoY) growth rates of imports to the U.S. from the top 20 exporting countries or territories. From Phase 1 to Phase 2, exports to the U.S. rose from the top 20 markets, except China. China had an average decline rate of -6.13% from Phase 1 to Phase 2 (Figure 2a), but the exports rebounded to 5.69% during the pandemic period from Phase 2 to Phase 3. Exports from other Asian markets except Japan continued to grow during this time. On the other hand, the other countries except for the Netherlands, Ireland, Russia, and Switzerland showed a declining trend in exports to the U.S. compared to the top Asian countries. In Europe, the exports to the U.S. from the Netherlands, Ireland, Russia, Switzerland, and Italy increased tremendously from Phase 1 to Phase 2 (Figure 2a). Specifically, the Netherlands exported 75.88% more goods to the U.S. during Phase 1 compared to Phase 2. Except for the United Kingdom and France, most European countries maintained or increased exports to the U.S. during the COVID-19 outbreak. In Phase 3, the export value of Switzerland increased by 53.09% compared to Phase 2. It is noteworthy that the Netherlands' growth rate declined substantially to 4.35%. The value of French exports to the U.S. declined by a significant -18.31%. A closer study of the volume variations across economic sectors indicates that the pandemic likely disrupted French exports and, presumably, its production of certain goods, such as medicine, medical devices, and equipment. Likewise, exports from the United Kingdom decreased by -15.96%.

From Phase 1 to Phase 2, the export value of the Top 20 countries and territories in North and South America to the U.S. increased slightly (Figure 2a). Nonetheless, exports from these three nations to the United States decreased throughout the epidemic, and Brazil's decline of -10.75% is the most severe.

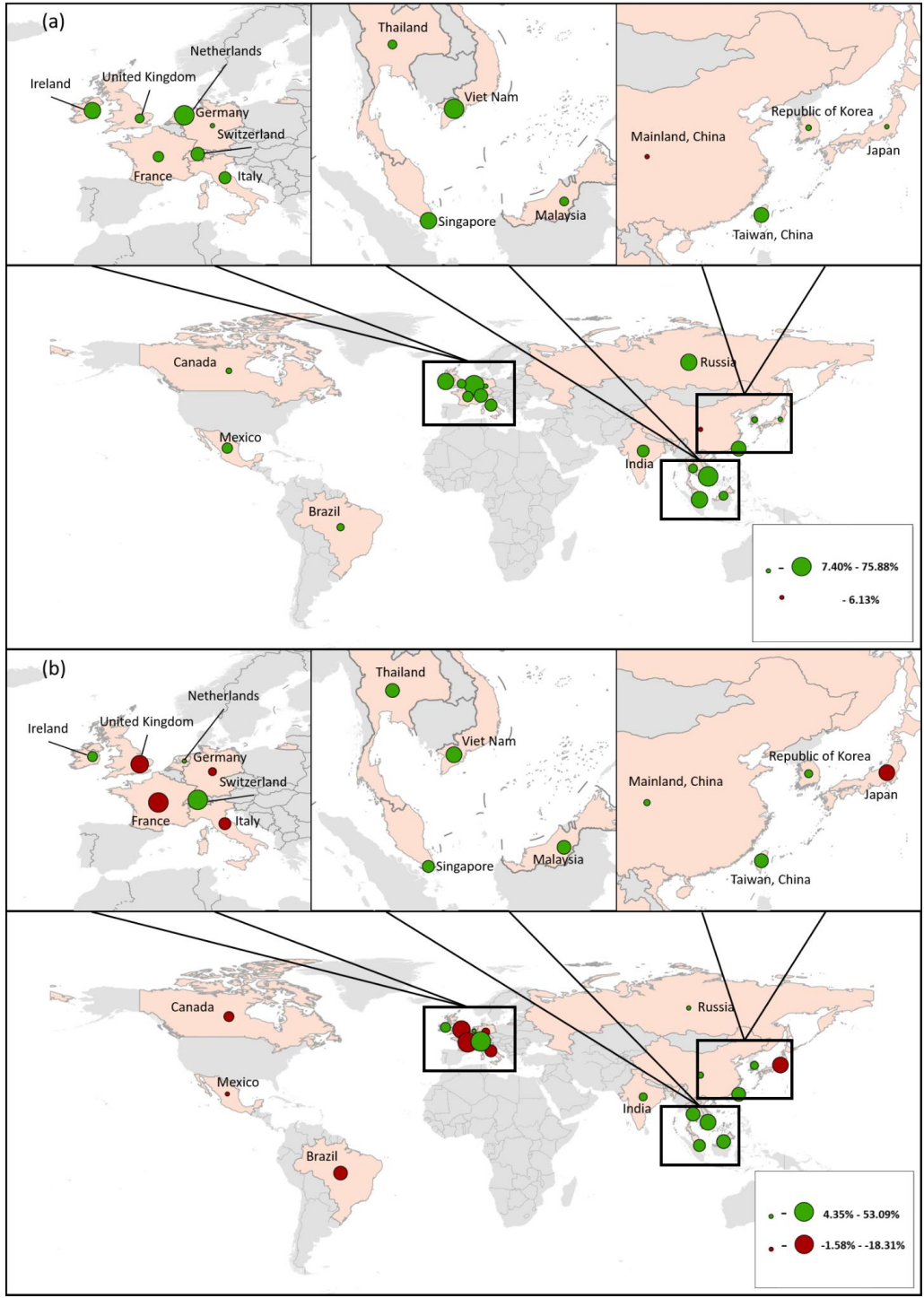

**Figure 2.** The Year-over-Year (YoY) growth rates of the U.S. imports from the top 20 countries and territories with the highest exports to the US. **(a)** The YoY growth rates of the U.S. imports between Phase 1 (2015-2017) and Phase 2 (2019). **(b)** The YoY growth rates of the U.S. imports between Phase 2 and Phase 3 (2020-2021). The green indicates positive growth between the two

phases, while the red indicates decline. The size of the circle means the value of the positive or negative growth rate.

## The U.S. Imports-Event-study Analysis

In this section, we use an event-study model to examine the effects of the trade war and the COVID-19 pandemic on U.S. imports. To explore the impacts of the trade war and epidemic, we categorize exporters to the U.S. into groups and investigate the differences in their trading activities across periods. We first split our samples from 2015 to 2021 into three Phases, i.e., 2015-2017, 2019, and 2020-2021, based on various scenarios. To measure the impacts of the trade war and pandemic, we assess the "norm" of U.S. import activities of each group with subsamples from 2015 to 2017, allowing us to control seasonality, business cycle, etc. To evaluate the effect of the trade war, we eliminate observations in 2018 to alleviate the short-term regulatory ambiguity and strategic procurement associated with tariff expectation or speculation. To measure the joint effects of the trade war and COVID-19, we use the same 'norm'. In contrast, to isolate the COVID-19 impact on U.S. imports, we employ measures generated from the 2019 sample to approximate the 'norm' of the trade war period. In an event study from analogous to standard difference-in-differences analysis, we separate our sample into focal and benchmark groups (equivalent to treatment and control groups) on the continent-sector and country-sector levels (Table S1 in the Supplementary Materials). In other words, the coefficient of estimates reflects the difference in percentage changes between the treatment and the control group in the period of interest compared to the period of normality. At the continent-sector level, each focal group represents the exporters from a specific economic sector in one continent, and the corresponding benchmark group is their counterpart from the same industry in the other continents. At the country-sector level, each focal group is the exporter of one country (or territory), and the corresponding group represents the exporters in all other countries (or territories).

Figures 3 and 4 illustrate the impact of the trade war and the pandemic on U.S. imports in all economic sectors at the both continent level and the country level. Figure 3a indicates that the trade war adversely affects imports from China to the U.S. but has a mixed impact on imports from other countries or territories. The Chinese apparel, machinery, and miscellaneous sectors experienced the most significant declines with magnitudes of -7.71, -6.89, and -5.04, respectively, which are also the economic sectors corresponding to the top goods imported by the U.S.[51]. The U.S. raised tariffs on Chinese goods from many economic sectors from 2018 to 2019, especially those from mechanical components, materials, and electrical equipment[52]. On the other hand, the trade war had a positive impact on U.S. apparel imports from Africa, Europe, and the rest of Asia (with values of 0.57, 0.89, and 6.03, respectively), as well as on machinery imports from Mexico (with a value of 1.081) and from Vietnam and Taiwan (with values of 1.654 and 1.239 respectively). In addition, there is a negative impact on the material imports from other American nations with a negative value of -1.36, and Canada with a negative value of -1.385. The trade war had the least detrimental effect on the U.S. imports of rare minerals from Mainland China compared to other economic sectors, which was likely owing to the strong dependence of the U.S. on rare materials from China[53].

When analyzing the COVID-19 pandemic, one must consider its interaction with the trade war. Figure 3b depicts the COVID-19 outbreak and the trade war on U.S. imports. In general, it exhibits similar tendencies as the trade war alone in Phase II (Figure 3a), with some noticeable differences. Over this period, U.S. imports from China were adversely impacted for most commodities, except for transportation imports, which increased with a value of 1.00. Compared to Phase II, Chinese exports of textiles, machinery, materials, and metals have decreased considerably more, while downward trends on chemical and miscellaneous imports have diminished. Meanwhile, imported goods of agriculture, apparel, machinery, materials, and miscellaneous sectors from other Asian regions grow significantly. These include apparel exports from Vietnam and India with magnitudes of 5.758 and 1.358, respectively, machinery exports from Malaysia, Korea, Taiwan, Thailand, and Vietnam with magnitudes of 0.566, 0.678, 2.405, 0.684, and 3.371, respectively, and materials exports from India, Malaysia, Thailand, and Vietnam with magnitudes of 0.701, 1.563, 1.138, and 2.111, respectively.

To separate the impact of COVID-19 from that of the trade war, we utilize metrics derived from the 2019 sample to determine the 'norm' for U.S. imports during the trade war. Figure 3c depicts the possible impact of COVID-19 on U.S. imports. COVID-19 caused the U.S. to increase imports from China on transportation and chemicals with values of 0.88 and 1.33, respectively, but led to a decrease in agriculture, metals, and minerals imports. When focusing on machinery imports, the COVID-19 pandemic resulted in a decline in imports from American and European continents and a rise in the rest of Asia (ROA). In conjunction with Figure 4, machinery imports from Canada, France, and Germany declined during the pandemic, while imports from Malaysia, Taiwan, Korea, Thailand, and Vietnam increased. This result implies that the quick economic recovery during the early stage of COVID-19 outbreak in China boosted machinery manufacturing activity in the downstream supply chain network near China and increased exports to the U.S., albeit not directly from Chinese exporters facing high tariffs. Conversely, machinery exports from other supply-chain clusters, such as those in Europe and the American continents, dropped dramatically presumably due to different epidemic management policies[54]. Furthermore, the COVID-19 outbreak had a remarkable positive impact on the U.S. importing materials from other American regions, metals from European areas, and materials and miscellaneous items from Asian regions.

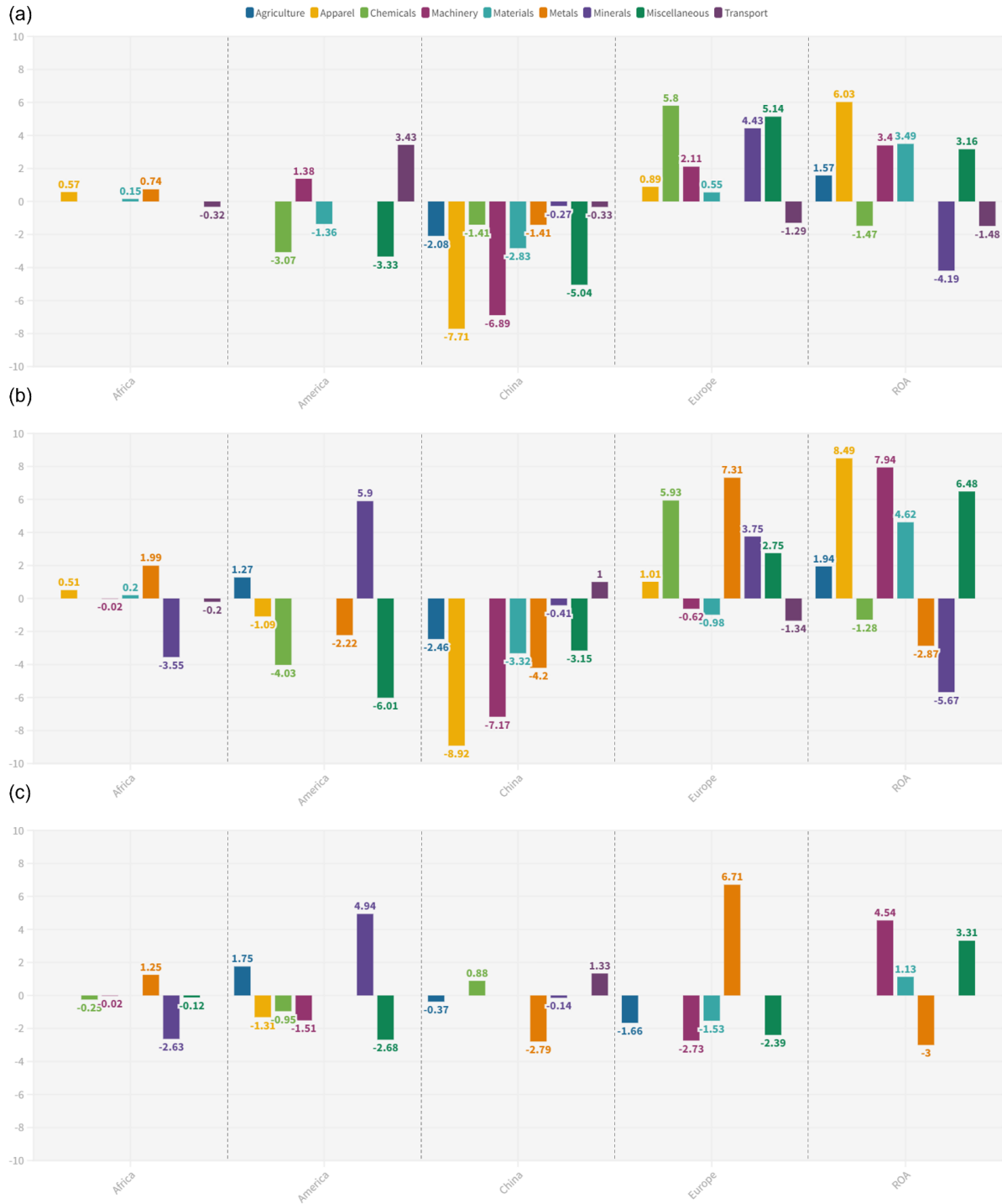

**Figure 3.** Event-study analysis of sector import shares (in percentage) by continent (or China) and sectors due to **(a)** the US-China trade war, **(b)** the US-China trade war and COVID-19 pandemic, **(c)** the COVID-19 pandemic. We benchmark each focal economic sector's monthly average market share from a continent (treatment) with all exporters outside that continent (control). We take 2015-2017 as the "normal" period (before), 2019 as the "post-event" era for (a), and 2020-2021 as the "post-event" period for (b). Moreover, we take 2019 as the "norm" period (before) and 2020-

2021 as the "post-event" era for (c). The resulting regression coefficients were assessed at 10% or lower significance levels to determine the most appropriate one for different data pairs.

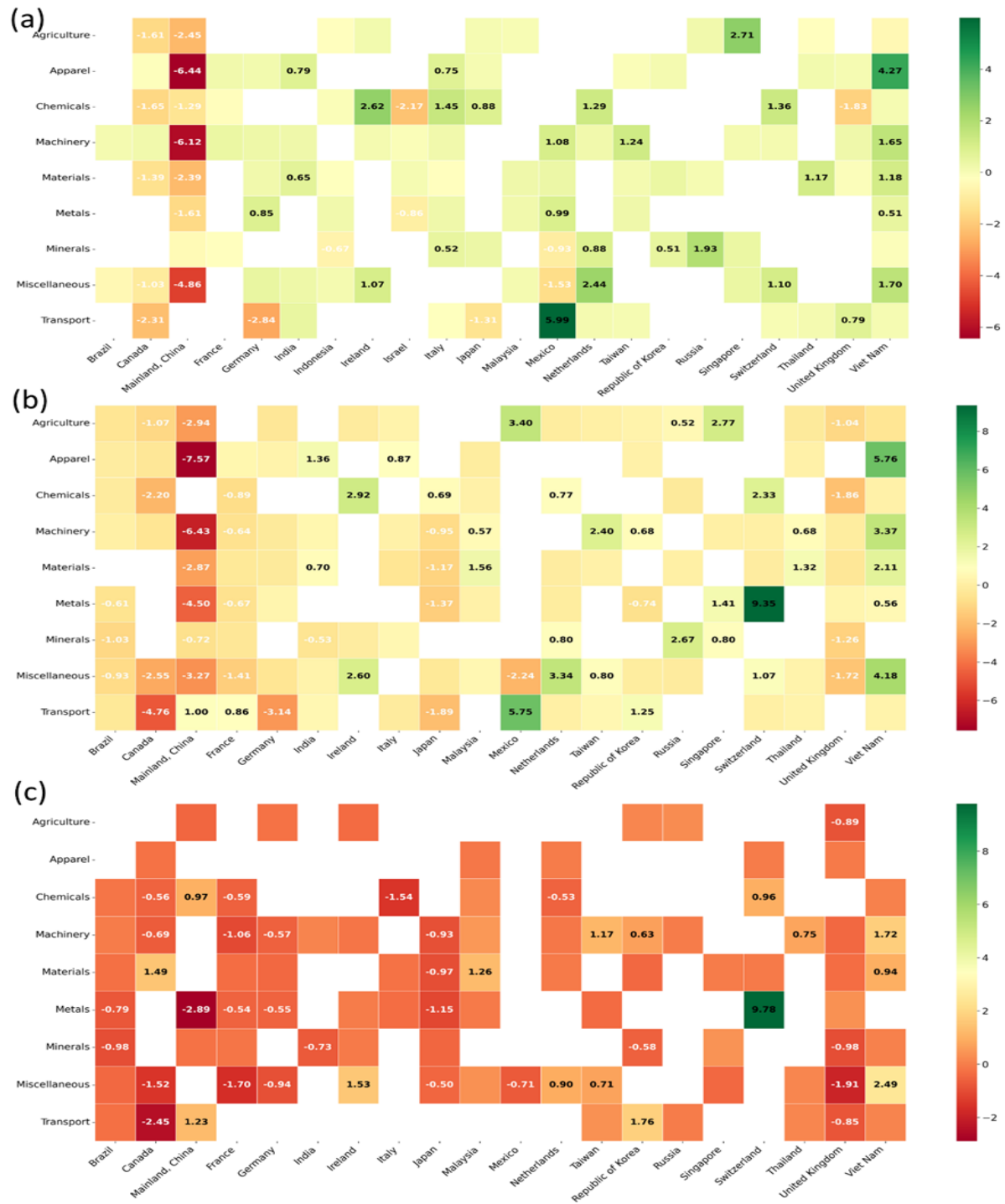

**Figure 4.** Event-study estimates at the country or territory level. The heat map shows the top 20 U.S. importers on the horizontal axis (in alphabetical order) and the nine economic sectors on the vertical axis. Each box corresponds to the impact of the event on that economic sector for that country or territory, with positive impacts marked in green and negative impacts in red, and the colors deepening as the impact increases. Values less than -0.5 and greater than 0.5 are labeled in this figure.

## China's Exports Analysis

Figure 5 illustrates a growth in China's exports to eight Asian countries (or territories) that were among the top 20 trade partners exporting to the U.S. In terms of value, China's export to these eight Southeast Asian countries is comparable to its export to the U.S. For example, in 2017, China's total export amount was $446 billion to those eight countries or territories and $563 billion to the U.S. In 2022, these figures become $642 billion and $541 billion, respectively. The substantial expansion of imports from China in this region suggests a growing dependency of the Asian supply chain on Chinese exporters. In combination with Figure 5a and Table 2a, the exports of China to these countries and territories exhibited a similar upward trend from 2016 to 2017. During Phase 2, China's exports to these Southeast Asian countries increased by 47.49%, 31.39%, 22.43%, and 20.40% for Vietnam, Malaysia, Singapore, and Thailand, respectively, compared to the rest of the countries and territories. Due to pandemic lockdowns, China's exports to these countries and territories declined significantly in Feb 2020 with a rapid growth rebound one month later.

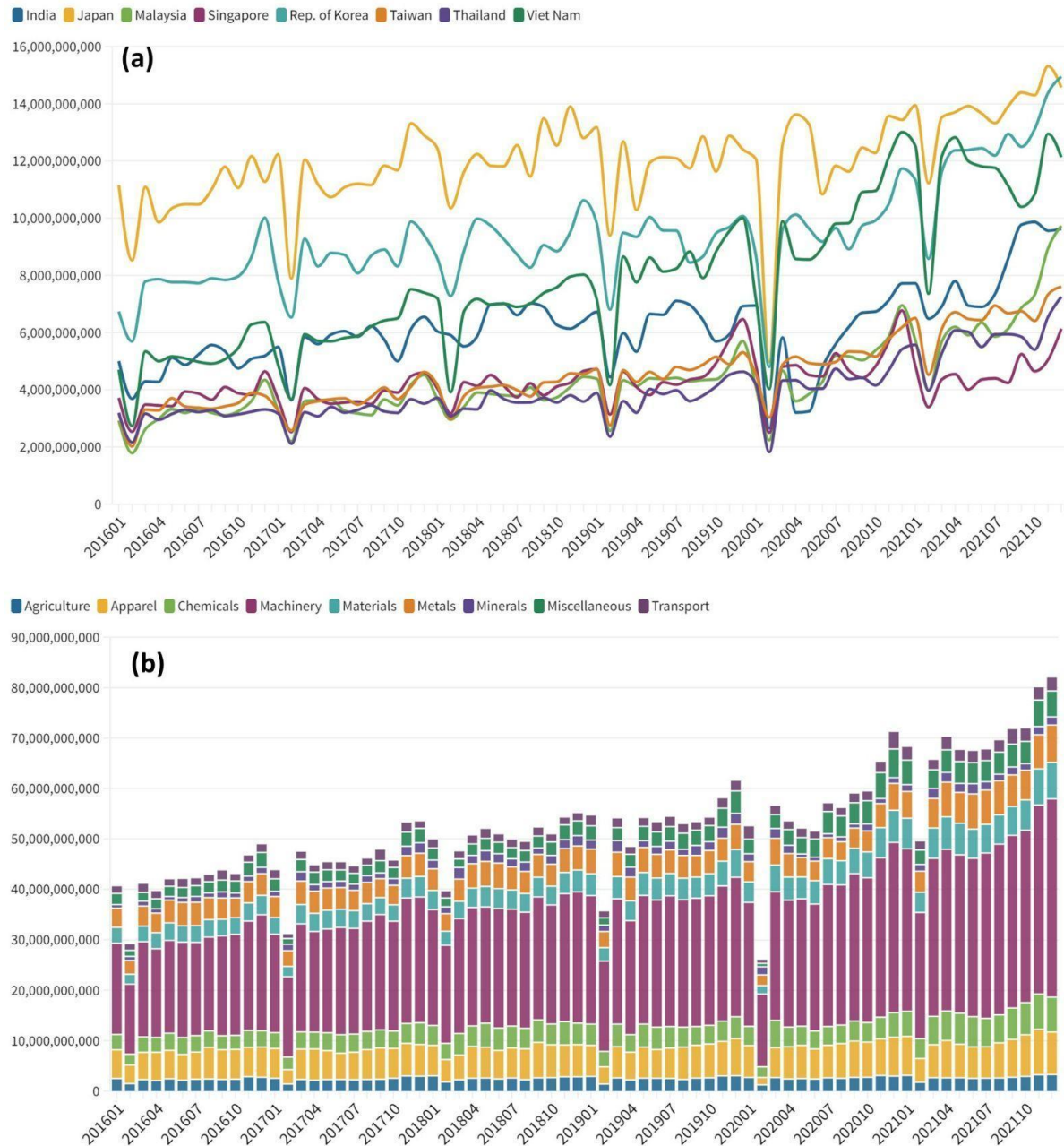

**Figure 5.** The monthly change in China's exports to eight Asian countries and territories that were among the top 20 trade partners exporting to the U.S. from 2016 to 2021. The Comtrade database does not include Chinese export records for 2015. Panel **(a)** displays monthly exports by countries and territories. Panel **(b)** illustrates the monthly exports at the economic sector level.

Table 2b reveals that China's export dynamics to these eight Asian countries or territories correspond to U.S. import dynamics in Table 1. For example, from Phase 1 to Phase 2, China's mineral exports increased the most by 48.21%, whereas the U.S. has the most significant increase in the same sector with 28.34%. Moreover, the exports of miscellaneous materials, equipment, and chemicals from China rose, while U.S. imports from the same economic sectors increased significantly. Between Phases 2 and 3, China's mineral exports decreased dramatically by -20.40%, mirroring the same changing pattern of U.S. mineral imports. Meanwhile, China exports more minerals, chemicals, and metals to these countries and territories, while the U.S. imports more goods from the same sectors.

**Table 2.** The annual change in China's exports to eight Asian countries or territories that were among the top 20 trade partners exporting to the U.S.. Phase 1, covering the period from 2016 to 2017, represents the pre-trade war period. Phase 2 is 2019, covering the trade war period. Phase 3 spans from 2020 to 2021, when the trade war and the COVID-19 pandemic clash.

**(a)**

|  | From phase1 to phase2 | From phase2 to phase3 |
|:---:|:---:|:---:|
| **India** | 18.36% | 9.74% |
| **Japan** | 7.49% | 7.65% |
| **Malaysia** | 31.39% | 29.45% |
| **Singapore** | 22.43% | 2.96% |
| **Rep. of Korea** | 13.00% | 17.74% |
| **Taiwan** | 30.98% | 25.70% |
| **Thailand** | 20.40% | 31.48% |
| **Viet Nam** | 47.49% | 28.60% |

(b)

|  | From phase1 to phase2 | From phase2 to phase3 |
|---|---|---|
| Agriculture | 7.60% | 4.14% |
| Apparel | 8.48% | 10.22% |
| Chemicals | 22.31% | 21.64% |
| Machinery | 23.50% | 19.74% |
| Materials | 28.26% | 24.57% |
| Metals | 10.81% | 13.96% |
| Minerals | 48.21% | -20.40% |
| Miscellaneous | 33.34% | 35.81% |
| Transport | 16.38% | 21.48% |

## China's Exports-Event-Study Analysis

During the same study period, we also do an event-study analysis to determine the influence of the combined effect of the trade war and the pandemic on China's exports to these eight Asian countries or territories. We employ the identical three baseline periods as we applied in the U.S.

imports event-study analysis, except that the first baseline period begins in 2016 due to the absence of 2015 China export data in the Comtrade database. Each treatment group represents a particular economic sector from one of these Asian countries and territories. In contrast, each control group represents the same sector of the remaining seven Asian countries and territories.

Figure 6a demonstrates that the commodities with the most significant decrease in China's export to the U.S. during the US-China trade war (Figure 3a) significantly increased in exports from China to developing Southeast Asian countries. In particular, we found that exports of apparel, machinery, and miscellaneous goods from China to Asian countries, such as Vietnam, Malaysia, and Thailand, significantly increase. In the machinery, minerals, and textiles sectors, there is a substantial alignment between Vietnam's imports from China and its exports to the U.S. (Figure 4).

Figure 6b illustrates the combined impact of the trade war and COVID-19 on China's exports. China's exports of machinery and materials sectors are expanding rapidly in Malaysia, Thailand, and Vietnam from which the U.S. imports significantly more. For instance, machinery exports from China to Vietnam and Malaysia had positive and significant coefficient estimates of 8.04 and 1.94, respectively. In contrast, its material exports to Vietnam, Thailand, and Malaysia had positive coefficient estimates of 7.04, 2.08, and 1.8, respectively.

Figure 6c depicts the isolated impact of the COVID-19 pandemic on China's exports to these eight Asian countries or territories. It indicates that China's exports of chemicals to the U.S., Thailand, and Vietnam increased in combination with Figure 3c, suggesting that Chinese suppliers play a crucial role in the worldwide combat against COVID-19. Over the same period, the U.S. increased machinery imports from ROA to which China's machinery exports differed. China's machinery exports to Japan and the Republic of Korea declined dramatically, whereas the exports to Thailand, Vietnam, and Taiwan grew significantly along their increased demand for medical device production[55].

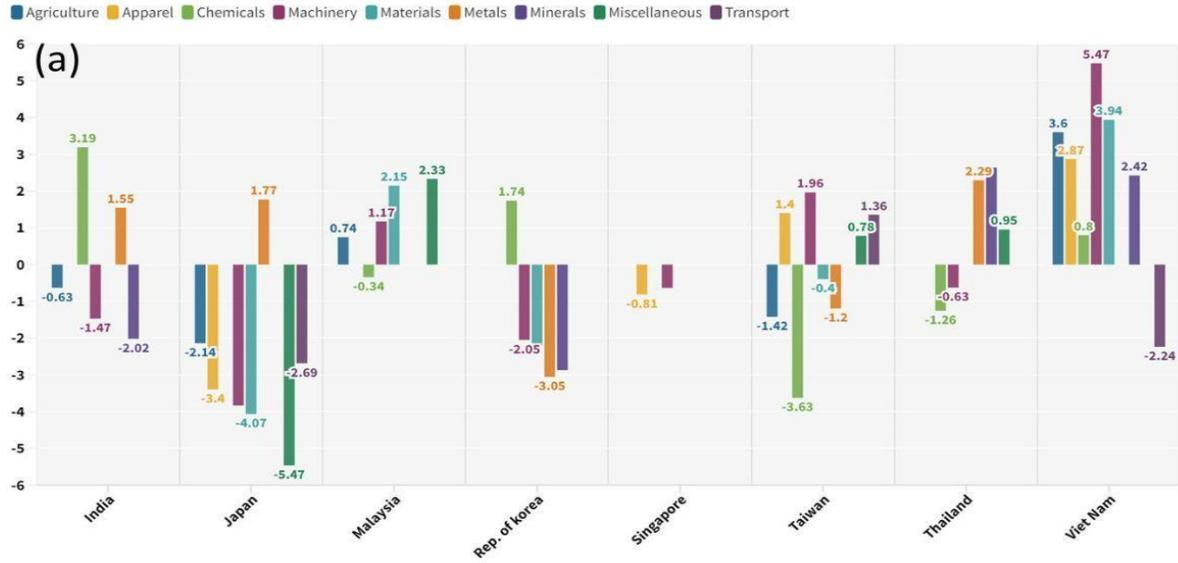
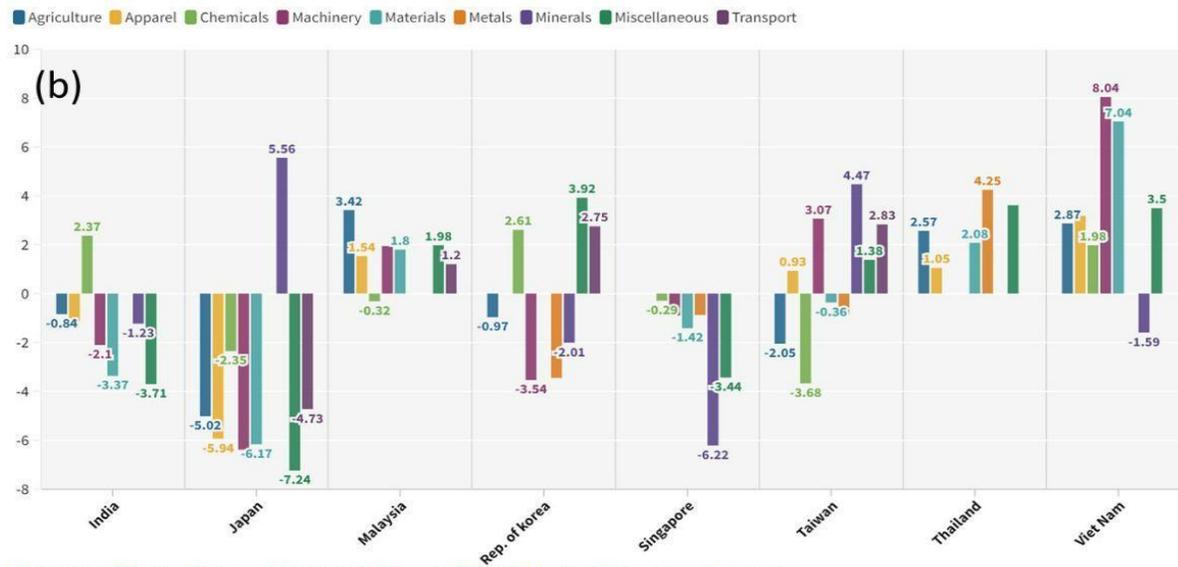
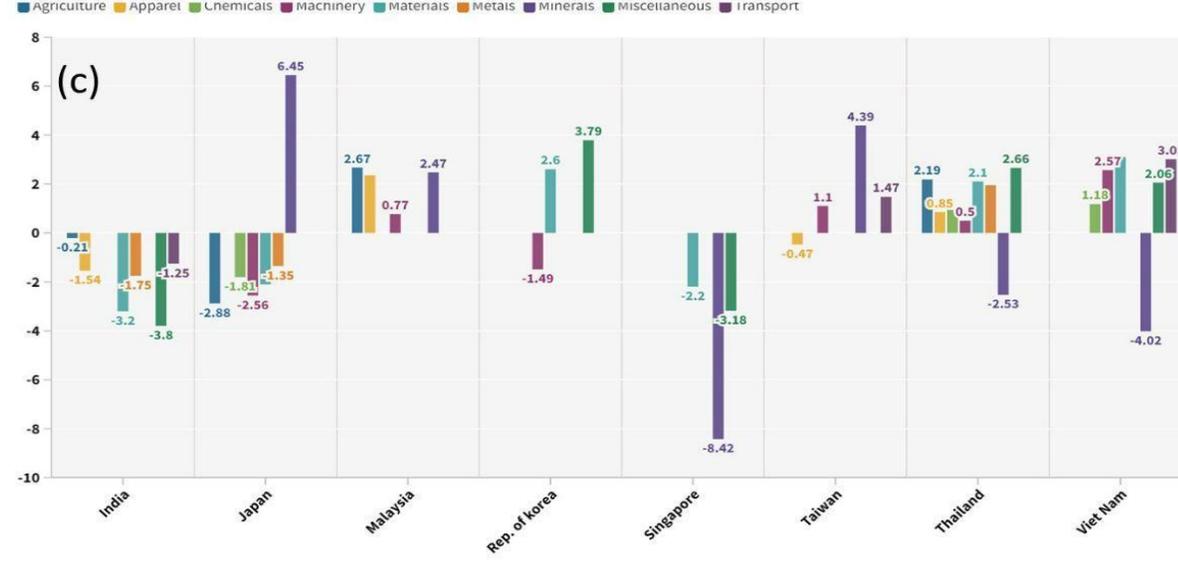

**Figure 6**. Event-study analysis of sector export shares (in percentage) by 8 Asian trading partners due to (a) the US-China trade war, (b) the US-China trade war and COVID-19 pandemic, (c) the COVID-19 pandemic. We benchmark each focal economic sector's monthly average export share to a country/territory (treatment) with all importers outside that country/territory (control). We take 2016-2017 as the "normal" period (before), 2019 as the "post-event" era for (a), and 2020-2021 as the "post-event" period for (b). Moreover, we take 2019 as the "norm" period (before) and 2020-2021 as the "post-event" era for (c). The resulting regression coefficients were assessed at 10% or lower significance levels to determine the most appropriate one for different data pairs.

## Discussion

This study demonstrates that the United States' imposition of increased tariffs had a significant impact on China's exports to the United States during the early stage of the trade war before the COVID-19 outbreak, forcing China to relinquish its market share to other countries. This transfer of market share exhibited heterogeneity in geographic dimensions. Some southeast emerging economies, such as Vietnam, Malaysia, and Thailand, benefited the most from China's diversion of exports away from the United States, while the European Union also potentially gained a modest advantage from the trade conflict[8,56–60]. The emerging economies of Southeast Asia (i.e., Vietnam, Malaysia, Thailand) have not only gained market share that originally belonged to China, but also taken over part of the value chain transferred from China.[17,61–64]. Our empirical study confirms this observation, showing that both the European Union and emerging economies of Southeast Asia (i.e., Vietnam, Malaysia, Thailand) experienced an increase in exports to the U.S. In addition, our study reveals that emerging economies of Southeast Asia also had surging imports from China that used to belong to the most negatively affected economic sectors (i.e., apparel, machinery, miscellaneous) from China to the U.S.. This suggests that Southeast Asia has undeniably become more integrated into the global supply chain system in segments traditionally dominated by Chinese suppliers.

Our research further shows that countries that had previously benefited from the effects of the trade war between the U.S. and China exhibited spatiotemporally heterogeneous challenges when facing the outbreak of COVID-19. For example, most Asian countries demonstrated more resilience than European and American countries in terms of their exports to the U.S.. Previous research also shows that China led the Asia-pacific countries to secure the global supply chain networks because of their disease control effectiveness[65,66]. Our research affirms the exceptional resilience of China's supply chain, enabling the country/territory to swiftly recover from the impact of the COVID-19 pandemic. Additionally, we find that the negative consequences of the US-China trade war have been significantly diminished during this period. This can be attributed to China's proactive implementation of stringent prevention and control policies in the early stages of the pandemic, as well as its well-developed industry chain system. As a result, Chinese exporters have gained a competitive advantage in sustaining continuous production following the COVID-19 outbreak. This has limited the options available to US importers, as the pandemic has continuously disrupted global supply networks outside of China along the COVID-19 waves, deterring value chains from relocating out of China.

Our analysis uncovers a growing triangular trade relationship involving China, Southeast Asia, and the United States, along with the trade war and COVID-19. Several key factors contribute to this trend. Firstly, the existing role of China as a pivotal hub within the Southeast Asian and East Asian trade bloc, connected through cross-border value chains, predates the trade conflict[67]. Secondly, the significant cost advantage derived from concentrated value chains in China presents challenges for upstream and downstream companies seeking to relocate[68]. Thirdly, after decades of globalization relying on comparative advantages, the complexity of reallocating production capacities and skilled labor from China to other nations poses a substantial burden[9]. Lastly, the economic strain on the United States and the costs associated with relocating value chains act as barriers to their migration from China[69]. Our findings indicate that, under the influence of the trade war, sectors such as apparel, machinery, and miscellaneous goods experience a decline in exports to the United States, which corresponds to an increase in exports of these sectors to Southeast Asia. This suggests the formation of a triangular trade relationship, wherein companies opt to transfer assembly processes to Southeast Asia to avoid additional U.S. tariffs. As a result, businesses can achieve cost savings by importing semi-finished goods from China, conducting assembly in non-China regions, and subsequently exporting to the United States. The COVID-19 reinforced the triangular trade relationships because the U.S. had to increase the import directly from China and indirectly through Southeast Asian countries that imported from China.

In conclusion, the US-China trade war has had substantial implications for the global supply chain and the subsequent COVID-19 pandemic has further reshaped the changing global supply chain. The escalated trade tension has shifted some segments associated with certain Chinese products and deepened the decoupling between the two world's largest economic bodies. However, the trade war initiated by the U.S. may not accomplish the objective of benefiting the U.S. businesses, especially those depending on overseas suppliers, and bypassing the cost of tariffs to their Chinese suppliers. The existing literature investigates the tariff mechanisms and calibrates the magnitude of tariff pass-through within economic models[42,46] and finds that the trade war is not harmless to the U.S. economy even in regular conditions. In contrast, some parties on the value chain network may have adopted a tariff avoidance strategy by forming the triangle trade framework. Furthermore, China's aggressive COVID policy helped its exporters and made some of them more formidable in the global economic arena, though at a substantial cost and with long-term uncertainty. There is no doubt, however, that the ongoing trade war results in higher input costs for U.S. importers and, subsequently, higher prices for U.S. consumers, especially in the early stage of the pandemic breakout. Even though this trade war's long-term consequence is mainly unknown, we hope this research keeps policymakers and the general public abreast of these measurable economic effects associated with the trade war.

Even though our study contributes valuable insights, we admit certain limitations and suggest further research directions. Future research can expand on our work by including U.S. exports to China, thereby providing a better understanding of the impacts of retaliatory tariffs on the downstream segments of the global supply chain, especially in the high-tech sector. In addition, future research might examine the changes in the consequential effects of the trade war in the aftermath of the pandemic, especially how various entities beyond two conflicting countries in global supply networks respond to the dual causes. Also, the lack of publicly accessible data prevents us from exploring the long-term impact of Chinese COVID-19 policies on the global supply chain beyond 2022.

## Description of supplemental information

Supplemental information includes two tables (Table S1 - Table S2) and seven Figures (Figure S1-S7).

## Declaration of Interests

**Competing Interest Statement:** The authors declare no competing interests.

## Acknowledgments

**Funding:** This work was supported in part by National University of Singapore FY2020 START-UP GRANT under WBS A-0003623-00-00.